\documentclass[aps,prb,showpacs,twoside,twocolumn,10pt]{revtex4-1}
\usepackage{times,epsfig,amssymb,amsfonts,amsmath, bm, subfigure,mathtools,amsthm,braket, soul,enumitem,xcolor, physics, graphics,graphicx, float}
\UseRawInputEncoding
\usepackage[normalem]{ulem}
\usepackage{comment}
\usepackage{epstopdf}

\usepackage{multirow}
\usepackage[colorlinks=true, citecolor=red, urlcolor=blue ]{hyperref}

\begin{document}

\title{Interfacial Control of both Magnetism and Polarization in a van der Waals Ferromagnet/Ferroelectric Heterostructure}

\author{Priyanshu Raj$^{1}$, Sourav Mal$^{2,3}$, Rana Saha$^{1}$, Prasenjit Sen$^{2,3}$}
\affiliation{$^1$Indian Institute of Science Education and Research (IISER), Tirupati 517507, India}
\affiliation{$^2$Harish-Chandra Research Institute, Chhatnag Road, Jhunsi, Prayagraj 211019, India}
\affiliation{$^3$Homi
Bhabha National Institute, Training School Complex, Anushakti Nagar, Mumbai 400094, India}

\begin{abstract}
Two-dimensional multiferroic van der Waals heterostructures provide a promising platform for the simultaneous control of distinct ferroic orders, with potential applications in magnetoelectric devices and spintronics. The practical implementation of such technologies requires 2D magnets with high Curie temperatures and strong perpendicular magnetic anisotropy (PMA). Here, based on first-principles calculations, we propose a multiferroic heterostructure composed of the room-temperature ferromagnet $\text{Fe}_3\text{Ga}\text{Te}_2$ and the ferroelectric $\text{In}_2\text{Se}_3$. We show that intercalation of Fe atoms into the van der Waals gap of the $\text{Fe}_3\text{Ga}\text{Te}_2$/$\text{In}_2\text{Se}_3$ heterostructure enhances PMA by nearly an order of magnitude relative to the pristine $\text{Fe}_3\text{Ga}\text{Te}_2$ monolayer, while simultaneously allowing electric polarization to be modulated through interfacial charge redistribution. The enhancement of PMA arises from interfacial hybridization that modifies the spin-orbit coupling of Fe $d$-orbitals. Our results demonstrate an effective pathway to engineer magnetoelectric coupling in two-dimensional multiferroic heterostructures and pave the way toward energy-efficient spintronic devices.
\end{abstract}

\maketitle


\section{Introduction}
The breakdown of scaling laws~\cite{Moore} has contributed to the stagnation of conventional CMOS performance~\cite{Solomon}, prompting the search for alternative, energy-efficient approaches based on precise control of electronic and magnetic properties~\cite{Fert, Manipatruni2018-gd}. In magnetic materials, the tunability of perpendicular magnetic anisotropy (PMA) is crucial for the development of next-generation spintronic devices. Notably, the presence of strong PMA is essential for stabilizing long-range magnetic order in 2D systems against thermal fluctuations, which would otherwise suppress such order in accordance with the Mermin-Wagner theorem~\cite{mermin-wagner}. High PMA also ensures low switching energy and low write and error rate~\cite{Wang}.
Over the past decade, this has spurred an intensive search for novel materials that enable breakthroughs in device physics. Technologies such as magnetic random access memory (MRAM)~\cite{Tang1995} and spin transfer torque (STT-RAM)~\cite{Taniguchi, Slonczewski1989-kx}, spin orbit torque (SOT-RAM)\cite{Nguyen2024-ov} exploit spintronics principles, using electron spin for nonvolatile and energy-efficient memory operations. However, current spintronic devices rely on electrical currents to manipulate spin\cite{Nozaki2019-qr}, resulting in higher energy consumption. Memory devices that enable fully electrical read and write operations \cite{Kent2015-lj} hold promise for faster and more energy-effective data storage. The emergence of  2D van der Waals (vdW) magnets has sparked a revolution in this field owing to their unique advantages, such as tunability and flexibility, which facilitate their integration into multilayer heterostructures.
The discovery of spontaneous magnetic order in thin films of \text{Cr}$_2$\text{Ge}$_2$\text{Te}$_6$ (CGT)~\cite{Gong} and \text{Cr}\text{I}$_3$\cite{Huang} along with the realization of room-temperature ferroelectrics such as  \text{In}$_2$\text{Se}$_3$\cite{Zhou} has opened new avenues for exploring coupled ferroic orders in their heterostructures. The interaction between layers can introduce new functionalities that are not present in individual layers. They show significant modification in electronic and magnetic properties through interlayer orbital hybridization~\cite{Burch}. 

Interfacial coupling in multiferroic heterostructures composed of two ferroic materials enables robust and reversible control over magnetic and ferroelectric properties, which is vital for the advancement of ultralow-power, nonvolatile logic and memory technologies.
Liang et al.~\cite{liang} demonstrated electric control of magnetism in a CGT/P(VDF-TrFE) heterostructure, where a small gate voltage of 5V toggles the magnetic hysteresis loop. This effect is attributed to the polarization-dependent interfacial hybridization between the ferroelectric polymer and CGT. Gong et al.~\cite{Gong2019-ct} demonstrated that reversing the polarization of In\textsubscript{2}Se\textsubscript{3} switches the magnetic anisotropy of CGT, highlighting strong magnetoelectric effects at the interface. Voltage switching of magnetic anisotropy is achieved in \text{Fe}$_3$\text{I}$_2$/Ir~\cite{Sun} depending on Ir thickness. However, the low Curie temperature of most known 2D magnets limits their practical application. The experimental realization of high-temperature ferromagnetism (FM) in 2D materials has gained recent attention. In particular, the monolayer \text{Fe}$_3$\text{Ge}\text{Te}$_2$ has been shown to exhibit a Curie temperature ($T_c$) of approximately 130~K\cite{Fei}, which can be further enhanced up to 300~K through electrostatic gating~\cite{Deng}. Another structurally analogous compound, \text{Fe}$_4$\text{Ge}\text{Te}$_2$~\cite{Seo}, demonstrates robust ferromagnetism with a $T_c$ of around 270~K. 

In this study, we focus on a related compound, \text{Fe}$_3$\text{Ga}\text{Te}$_2$ (FGaT)~\cite{Zhang}, for which above room temperature $T_c$ in the range of 350-380 K has been reported for few-layer samples~\cite{Zhang}, although with relatively weak perpendicular magnetic anisotropy. These compounds share a similar structure characterized by a quasi-three-dimensional network~\cite{Seo} of Fe atoms, which enhances the exchange interaction~\cite{Ruiz} and is likely responsible for their elevated Curie temperatures. Numerous studies focus on control of magnetism in these high $T_c$ 2D magnets~\cite{Wang2023-pp, Wang2023-us, Eom2023-as}. 

Particularly interesting behavior has been observed in heterostructures made of FM and ferroelectric (FE) layers. In a heterostructure comprising FM \text{V}\text{Bi}$_2$\text{Te}$_4$ and FE \text{In}$_2$\text{Se}$_3$ monolayers\cite{Wang2023-pp}, magnetic anisotropy in \text{V}\text{Bi}$_2$\text{Te}$_4$ can be reversibly switched between in-plane and out-of-plane orientations by controlling the polarization state of the FE layer. This transition is accompanied by a semiconductor-to-half-metal phase change, driven by band alignment and interfacial charge transfer. In related developments, spin-orbit torque (SOT) switching has been demonstrated in \text{Fe}$_3$\text{Ge}\text{Te}$_2$/\text{Bi}$_2$\text{Te}$_3$\cite{Wang2023-us}, noted by enhanced Curie temperature of \text{Fe}$_3$\text{Ge}\text{Te}$_2$ and serves as an efficient spin current source, enabling low current magnetization switching. Furthermore, application of voltage in the  \text{Fe}$_3$\text{Ge}\text{Te}$_{3-x}$/\text{In}$_2$\text{Se}$_3$~\cite{Eom2023-as}  heterostructure leads to a reduction of coercive fields irrespective of voltage polarity. This result is attributed to the in-plane tensile strain as confirmed by DFT calculation of magnetic anisotropy. These findings collectively highlight the potential of vdW multiferroic heterostructures as platforms for next-generation spintronics and memory applications.

In this study, through first-principles density functional theory (DFT) calculations based on the 2D heterostructure of ML \text{Fe}$_3$\text{Ga}\text{Te}$_2$ and \text{In}$_2$\text{Se}$_3$, we observe a huge increase in magnetic anisotropy in the FM layer for both the polarization states of ferroelectric \text{In}$_2$\text{Se}$_3$. In the heterostructure, we find an increase in magnetic anisotropy as compared to the pristine monolayer \text{Fe}$_3$\text{Ga}\text{Te}$_2$ along with an increase in the magnitude of dipole moment in the direction of up polarization. Moreover, magnetic anisotropy and dipole moment increase even more when Fe atoms are intercalated in the interlayer region between \text{Fe}$_3$\text{Ga}\text{Te}$_2$ and \text{In}$_2$\text{Se}$_3$. These results demonstrate that a \text{Fe}$_3$\text{Ga}\text{Te}$_2$/\text{In}$_2$\text{Se}$_3$ show multiferroic behavior with ferromagnetism with strong PMA and switchable electric polarization. This study highlights how interfacial coupling enables the tunability of magnetic properties, offering a promising pathway for next-generation spintronics devices.

\begin{table*}[htbp]
    \centering
    \caption{Comparison of vdW-DF and LDA functionals for obtaining correct structural and magnetic properties of bulk and monolayer FGaT.}
    \vspace{0.2cm}
    \begin{tabular}{|c|c|c|c|c|c|c|}
        \hline
        Method & System & $a = b$ (\AA) & $c$ (\AA) & Interlayer Separation (\AA) & $M$ ($\mu_B$/Fe) & MAE (meV/Fe) \\
        \hline
        {Experiment \cite{Zhang}} & Bulk & 3.986 & 16.229 & 7.8 & 1.68 & 0.111 \\
                                                
        \hline
        \multirow{2}{*}{PBE + rVV10L}            & Bulk & 3.993 & 15.655 & 7.827 & 1.91 & 0.443 \\
                                                 & ML   & 3.992 & 15.613 & *     & 1.92 & 0.088 \\
        \hline
        \multirow{2}{*}{LDA}                     & Bulk & 3.944 & 15.730 & 7.8   & 1.72 & 0.183 \\
                                                 & ML   & *     & 16.751 & *     & 1.79 & 0.253 \\
        \hline
    \end{tabular}
    \label{LDA vs vdW}
\end{table*}

\section{Computational Details}

We performed spin-polarized DFT calculations using the 
Vienna \textit{ab initio} Simulation Package (VASP) \cite{PhysRevB.54.11169}. We employed the projector augmented wave method \cite{PhysRevB.50.17953} to treat electron-ion interactions. The PBE+rVV10L\cite{PhysRevB.82.081101} functional was used to calculate the exchange-correlation energy
in the presence of van der Waals (vdW) interactions,
which has been shown to accurately capture both structural and magnetic properties. For benchmarking, additional calculations were performed using the local density approximation (LDA), and other non-local vdW functionals. Consistency of each method with experimental results was checked.
For bulk  FGaT, a $\Gamma$-centered \textit{k}-point mesh of size $15 \times 15 \times 3$ was used for Brillouin zone integration in self-consistent calculations with a plane-wave energy cutoff of 500 eV. We performed full structural relaxations, allowing both lattice parameters and atomic positions to vary until the total energy and atomic forces converged to below $10^{-6}$ eV and $0.01$ eV$/\text{\AA}$, respectively. 


For the heterostructures, a plane-wave energy cutoff of 450 eV, and a $\Gamma$-centered \textit{k}-point mesh of size $15 \times 15 \times 1$ were used for self-consistent calculations. A sufficiently large vacuum spacing ($ > 15$~\AA) was included to eliminate interactions between periodic images along the out-of-plane direction. 

For the calculation of magnetic anisotropy, we performed fully relativistic calculations by including spin-orbit coupling (SOC) in the Hamiltonian in a non-self-consistent manner. We calculated the total energies of the system by orienting the magnetization direction along the crystallographic a-axis (in-plane) and c-axis (perpendicular to the planes), yielding energies $E_a$ and $E_c$, respectively. The magnetic anisotropy energy (MAE) was then calculated as the difference between these energies:
\begin{equation}
    \text{MAE} = E_a-E_c
\end{equation}

Positive MAE ($E_a > E_c$) represents perpendicular magnetic anisotropy (PMA), with Fe spins preferring to align out-of-plane. For MAE calculations, we used a strict energy convergence threshold of $10^{-8}$ eV, and a k-mesh of  $18 \times 18 \times 3$ for bulk FGaT, and $18 \times 18 \times 1$  for heterostructures and monolayers.

\section{RESULTS AND DISCUSSIONS}

\begin{figure*}[t]
    \centering
    \includegraphics[width=0.8\textwidth]{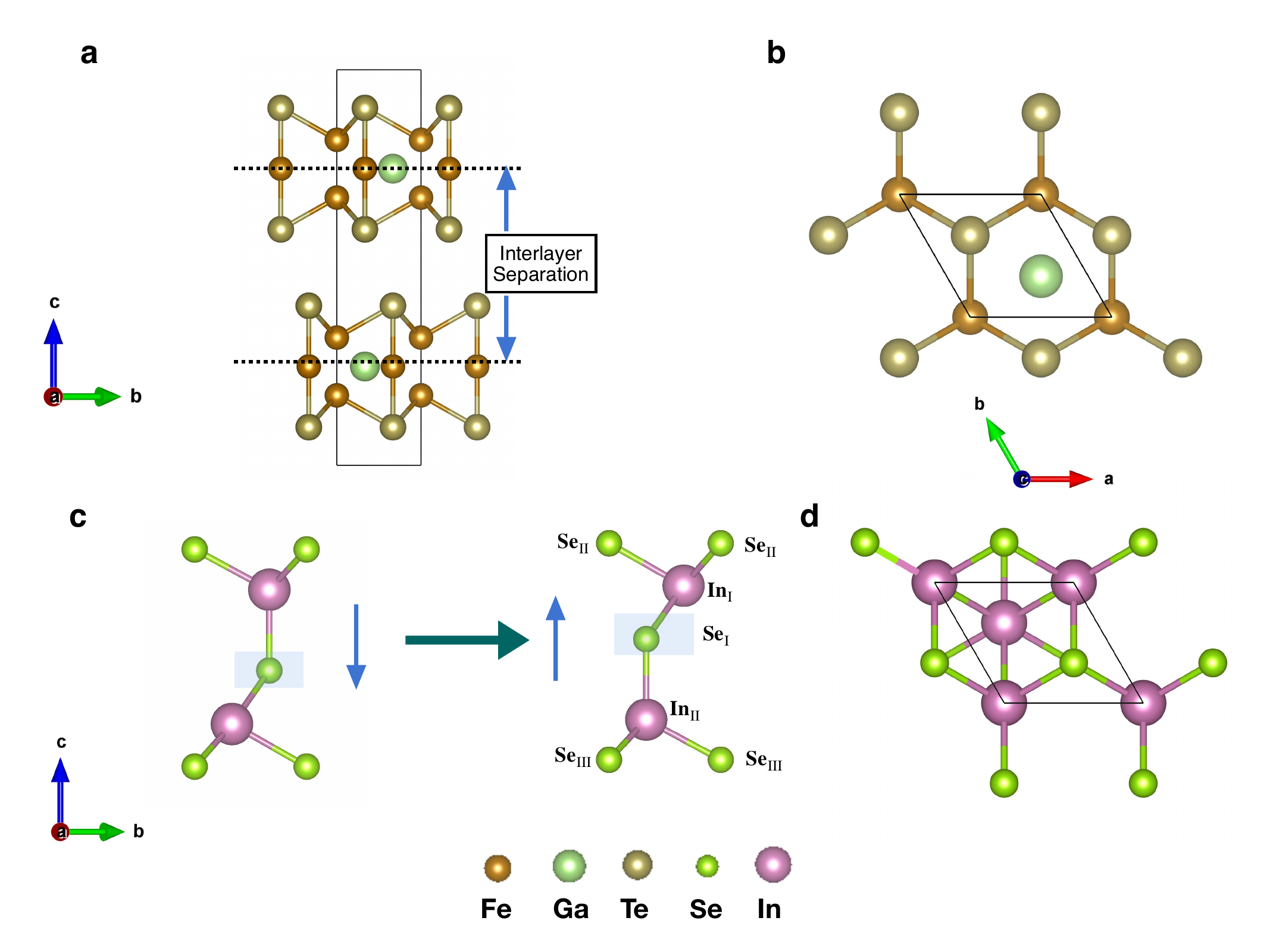}
    \caption{Visualization of the crystal structures (a) the side view of bulk FGaT, (b) the top view of bulk FGaT, (c) the side view of In$_2$Se$_3$ monolayer with up and down polarization directions. The middle Se layer is shaded in blue, (d) the top view of the In$_2$Se$_3$ monolayer.} 
    \label{crystal-structure-1}
\end{figure*}

\begin{figure*}[t]
    \centering
    \includegraphics[width=1\textwidth]{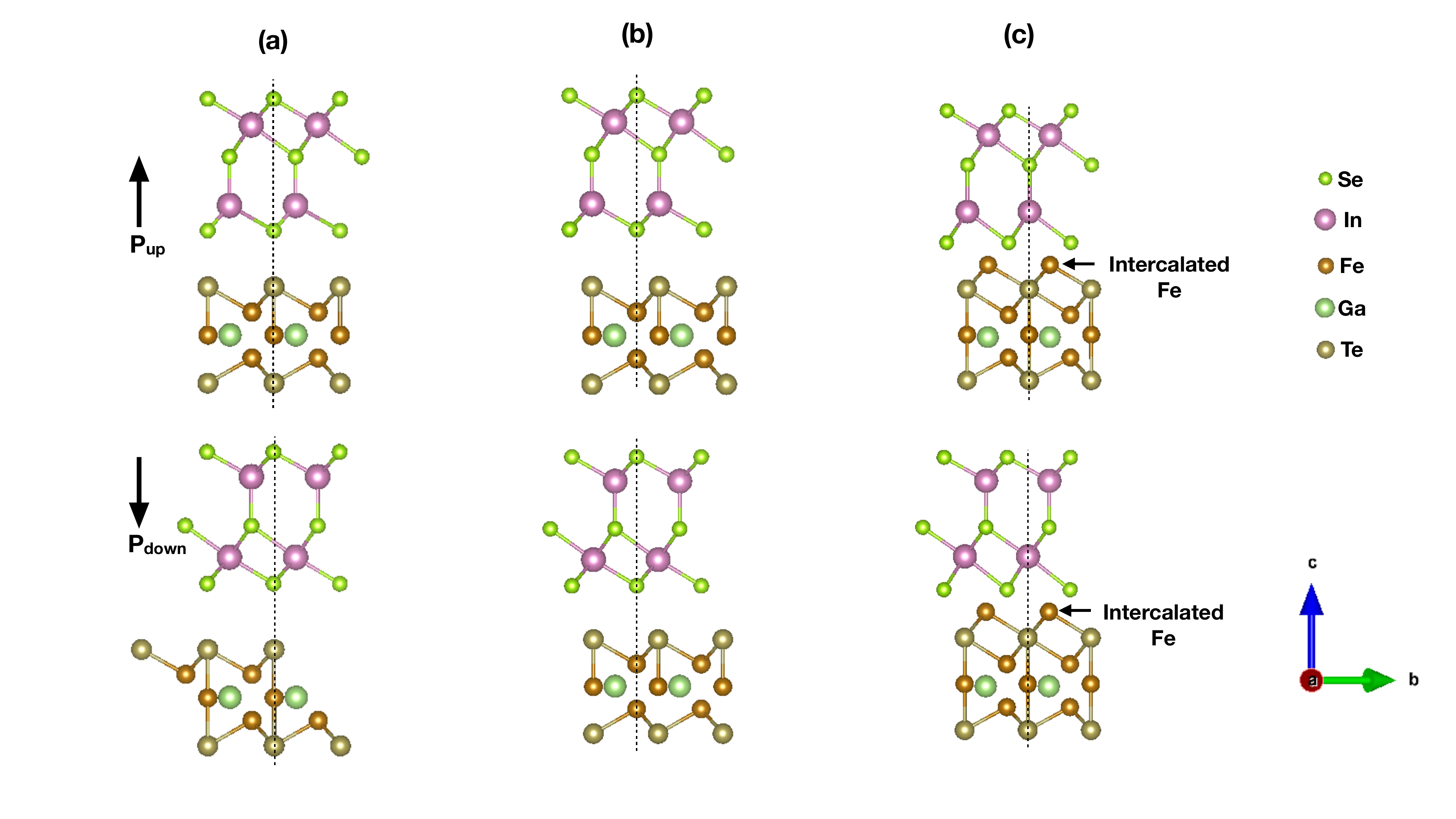}
    \caption{Visualization of the crystal structures of the $1 \times 1$ Fe$_3$GaTe$_2$/In$_2$Se$_3$ heterostructures. The polarization direction of In$_2$Se$_3$ is indicated by upward and downward arrows, representing the up and down polarization states, respectively. Dashed lines illustrate the alignment of interfacial Te and Se atoms for the hollow and top stacking configurations. 
(a) Fe$_3$GaTe$_2$/In$_2$Se$_3$ heterostructure with top stacking, 
(b) Fe$_3$GaTe$_2$/In$_2$Se$_3$ heterostructure with hollow stacking,
(c) Fe-intercalated Fe$_3$GaTe$_2$/In$_2$Se$_3$ heterostructure, with the intercalated Fe atom positioned at the hexagonal voids between the layers, as indicated by the arrow.} 
    \label{crystal-structure-2}
\end{figure*}

Bulk FGaT has a layered hexagonal structure with space group P63/mmc (No. 194). The experimental~\cite{Zhang} lattice parameters are $a = b = 3.9860\,\text{\AA}$, $c = 16.2290\,\text{\AA}$, $\alpha = \beta = 90^\circ$, $\gamma = 120^\circ$. It consists of two layers in a unit cell, as shown in Fig.~\ref{crystal-structure-1}(a). Each monolayer of FGaT consists of five sublayers where the Fe$_3$Ga slab is sandwiched between two Te atomic layers. The two monolayers are stacked along the c-axis with an interlayer separation of 7.8 $\text{\AA}$. The top view of the monolayer is shown in  Fig.~\ref{crystal-structure-1}(b).


To benchmark our results against experimental observations~\cite{Zhang}, we assess the validity of various exchange-correlation functionals in describing the properties of bulk FGaT, as summarized in Supplementary Information. 
While both LDA and the van der Waals functional PBE$+$rVV10L accurately reproduce the structural parameters and magnetic moments, LDA fails to capture the correct magnetic anisotropy. Specifically, it underestimates the bulk PMA, yielding a value of 0.183 meV/Fe, which is lower than the ML FGaT value of 0.253 meV/Fe. This is not surprising as LDA is not expected to work well for vdW materials. However, sometimes it leads to correct results by some lucky coincidence. As an example, it gives quite reasonable inter-layer separation in the case of graphite where GGAs fail~\cite{PhysRevB.46.7185}. In contrast, PBE$+$rVV10L shows better agreement with experimental data, predicting a lattice constant of 3.992~$\text{\AA}$ and an interlayer separation of 7.8~$\text{\AA}$. The bulk PMA value of 0.443 meV/Fe obtained with this functional is slightly overestimated, whereas the ML value of 0.088 meV/Fe is in close alignment with experimental measurements~\cite{Zhang}. A detailed comparison between LDA and PBE$+$rVV10L with respect to experimental values is provided in Table~\ref{LDA vs vdW}.

In$_2$Se$_3$ is a layered vdW ferroelectric (FE) material~\cite{Ding2017-ki} with out-of-plane polarisation. Monolayer In$_2$Se$_3$ consists of five atomic layers in the sequence of Se-In-Se-In-Se along the c-axis. Bulk In$_2$Se$_3$ exists in two phases, namely $\alpha$ and $\beta$~\cite{Osamura1966-rz} based on stacking of two monolayers. DFT supported calculations~\cite{Ding2017-ki} compared with experimental findings~\cite{Tao2013-vg} indicate $\alpha$ phase to be the ground state with in-plane lattice constants of  4.106 and 4.108 \text{\AA}, respectively. 
The side view and top views of monolayer In$_2$Se$_3$ are shown in Fig.~\ref{crystal-structure-1}(c)-(d). It exhibits spontaneous out-of-plane electric polarization, which can be reversed via laterally shifting the middle Se layer (highlighted by the blue shaded region in Fig.~\ref{crystal-structure-1}(c)). The two structures shown in Fig.~\ref{crystal-structure-1}(c) are energetically degenerate for an isolated ML with opposite polarizations along the c-axis, as shown in the figure, with reported electric dipole moment of magnitude $0.094$~e{\text{\AA}} per unit cell. 



The PBE$+$rVV10L functional tested above for ML FGaT is then further tested for ML \text{In}$_2$\text{Se}$_3$ where we obtain in-plane lattice constants of 3.985 $\text{\AA}$ and 3.986 $\text{\AA}$ for up and down polarization, respectively, with a magnitude of 0.096 e$\text{\AA}$ for the dipole moment close to the reported values \cite{Ding2017-ki}. We henceforth use PBE$+$rVV10L for the study of FGaT/\text{In}$_2$\text{Se}$_3$ heterostructures, unless otherwise stated.

The close lattice match enables the formation of FGaT/\text{In}$_2$\text{Se}$_3$ heterostructure with negligible strain.
We consider two possible vertical stacking configurations for the heterostructure. In top stacking, the interfacial Te atoms in FGaT sit on top of the In atoms in the In$_2$Se$_3$ layer. In the hollow stacking, the Te atoms sit above hollow sites created by the outer sub-layer In atoms. These are shown in Fig.~\ref{crystal-structure-2}(a) and (b), respectively. Once the In$_2$Se$_3$ forms a heterostructure
with FGaT, the two polarization directions are no longer symmetry-equivalent anymore. The two orientations of the electric dipole moment are designated as `up' and `down' as shown in Fig.~\ref{crystal-structure-2}. 
The hollow stacking configuration is energetically more favourable than the top stacking by more than 160 meV per unit cell for both polarization directions.


The equilibrium interlayer distance between FGaT and \text{In}$_2$\text{Se}$_3$ in the hollow-site configuration was found to be 2.93~\AA\ and 2.82~\AA\ for the up and down polarization states of In$_2$Se$_3$, respectively.  
Notably, the total energy of the downward-polarized configuration is lower by 55 meV/u.c. compared to the upward-polarized one.

In this heterostructure, we find significant modulation of PMA and polarization, suggesting interfacial coupling between the FM and FE layers. We note a striking enhancement of magnetic anisotropy. As shown in Table~\ref{pristine_het}, PMA is 0.210 and 0.225 meV/Fe for the up and down polarization states, respectively. These values are markedly higher than the PMA observed in an isolated ML FGaT. Additionally, the out-of-plane dipole moment in the In$_2$Se$_3$ layer for the `up' polarization direction is found to be $+0.25$, a value significantly larger than that in the isolated ML In$_2$Se$_3$. The polarization for the `down' direction, interestingly, increases from $-0.094$~eV\AA~ for an isolated ML to $-0.059$ eV\AA~. These changes indicate effective charge redistribution in the heterostructure. 
The magnetic moment per Fe atom remains nearly unchanged at ∼1.93$\mu_B$.

\begin{table*}[htbp]
    \centering
    \caption{Modulation of perpendicular magnetic anisotropy (PMA) and dipole moment in the Fe$_3$GaTe$_2$/In$_2$Se$_3$ heterostructure for different polarization states of In$_2$Se$_3$.}
    \vspace{0.2cm}
    \begin{tabular}{|c|c|c|c|c|c|}
    \hline
    Polarization & $a$ (\AA) & Dipole Moment (e\AA) & Interlayer Gap (\AA) & MAE (meV/Fe) & $M$ ($\mu_B$/Fe) \\
    \hline
    Up   & 3.988 &  0.25       & 2.88 & 0.210 & 1.927 \\
    Down & 3.985 & $-0.059$    & 2.84 & 0.225 & 1.935 \\
    \hline
    \end{tabular}
    \label{pristine_het}
\end{table*}

Further, we intercalate Fe atoms in this heterostructure in the ratio of one Fe atom per unit cell. Self-intercalation in the vdW gaps has emerged as a highly effective strategy for tailoring the magnetic and electronic properties of 2D vdW materials~\cite{Zhao2020-lg, Fujisawa}.
The intercalated Fe atoms are positioned at the centroid of the hexagonal voids formed between two opposing triangles of the Te and Se lattices.
Fe intercalated FGaT/\text{In}$_2$\text{Se}$_3$ heterostructure leads to a dramatic enhancement of both magnetic anisotropy and polarization. As presented in Table~\ref{Fe het}, the PMA increases to 0.591 meV/Fe for the up polarization state and 0.683 meV/Fe for the down polarization state. These values are more than double the PMA in the pristine heterostructure. Such a substantial increase reflects strong spin-orbit coupling modulation induced by hybridization between the intercalated Fe atom.
Furthermore, the total magnetic moment also shows a noticeable increase, reaching 2.16 and 2.24 $\mu_B$/Fe for up and down polarization states, respectively.
 Notably, the dipole moments reach 0.49 and 0.019 $e\text{\AA}$ for `up' and `down' polarizations, respectively. Note that the dipole moment increases further compared to the pristine heterostructure
 for the `up' configuration. For the `down' configuration also, the value increases, and in this case, flips to a positive value. Thus, the same structural distortion that produced a dipole moment directed towards 
 the interface (negative value) for the pristine heterostructure now produces a dipole moment away from the interface (positive value). 

To understand the origin of this polarization flip and its enhanced magnitude, we performed a detailed analysis of the local atomic structure in the FE layer. While the internal structure of the FE layer remains consistent between its two polarization states within a given system (pristine or intercalated), pronounced differences emerge between the pristine and Fe-intercalated cases. As summarized in Table~\ref{bond-FE}, for the up polarization state, the In--Se bond lengths adjacent to the FGaT layer exhibit a marked increase upon Fe intercalation. Furthermore, the bond angles undergo substantial distortion; for instance, the Se$_\mathrm{III}$--In$_\mathrm{II}$--Se$_\mathrm{III}$ angle decreases from $97.74^\circ$ to $87.73^\circ$, while the Se$_\mathrm{III}$--In$_\mathrm{II}$--Se$_\mathrm{I}$ angle increases from $119.56^\circ$ to $126.93^\circ$. These significant structural changes demonstrate that Fe intercalation induces a strong lattice distortion in the In$_2$Se$_3$ layer, which is the direct physical mechanism responsible for the enhanced and inverted polarization. A similar trend is observed for the down polarization state, confirming the general nature of this effect.

This again suggests considerable charge redistribution at the interface, and we analyze this below.
In any case, this tunability of magnetic and electric dipole properties by interface engineering alone is surely interesting and attractive.

\begin{table*}[htbp]
\centering
\caption{Bond lengths and bond angles in the $\text{In}_2\text{Se}_3$ layer for the up polarization state in pristine system, FGaT/$\text{In}_2\text{Se}_3$ heterostructure, and the Fe-intercaleted heterostructure.}
\vspace{0.2cm}
\begin{tabular}{|c|c|c|c|}
\hline
Quantity & Pristine & FGaT/$\text{In}_2\text{Se}_3$ & Fe-intercaleted FGaT/$\text{In}_2\text{Se}_3$ \\
\hline
In$_\mathrm{II}$-Se$_\mathrm{III}$ (\AA) & 2.65 & 2.65 & 2.88 \\ 
In$_\mathrm{II}$-Se$_\mathrm{I}$ (\AA) & 2.53 & 2.53 & 2.60 \\
In$_\mathrm{I}$-Se$_\mathrm{I}$ (\AA) & 2.84 & 2.84 & 2.83 \\
In$_\mathrm{I}$-Se$_\mathrm{II}$ (\AA) & 2.67 & 2.67 & 2.68 \\
$\angle$ Se$_\mathrm{III}$-In$_\mathrm{II}$-Se$_\mathrm{III}$ ($^\circ$) & 97.74 & 97.35 & 87.73  \\
$\angle$ Se$_\mathrm{III}$-In$_\mathrm{II}$-Se$_\mathrm{I}$ ($^\circ$) & 119.56 & 119.86 & 126.93 \\ 
\hline
\end{tabular}
\label{bond-FE}
\end{table*}

\begin{table*}[htbp]
\centering
\caption{Effects of Fe intercalation in interlayer hexagonal void of FGaT/\text{In}$_2$\text{Se}$_3$ heterostructure.}
\vspace{0.2cm}
\begin{tabular}{|c|c|c|c|}
\hline
Polarisation & Dipole Moment (e\AA) & MAE (meV/Fe) & M ($\mu_B$/Fe) \\
\hline
Up   & 0.49 & 0.591 & 2.16     \\
Down & 0.019 & 0.683  & 2.24   \\
\hline
\end{tabular}
\label{Fe het}
\end{table*}

\begin{figure}[htbp]
    \centering
    \includegraphics[width=0.5\textwidth]{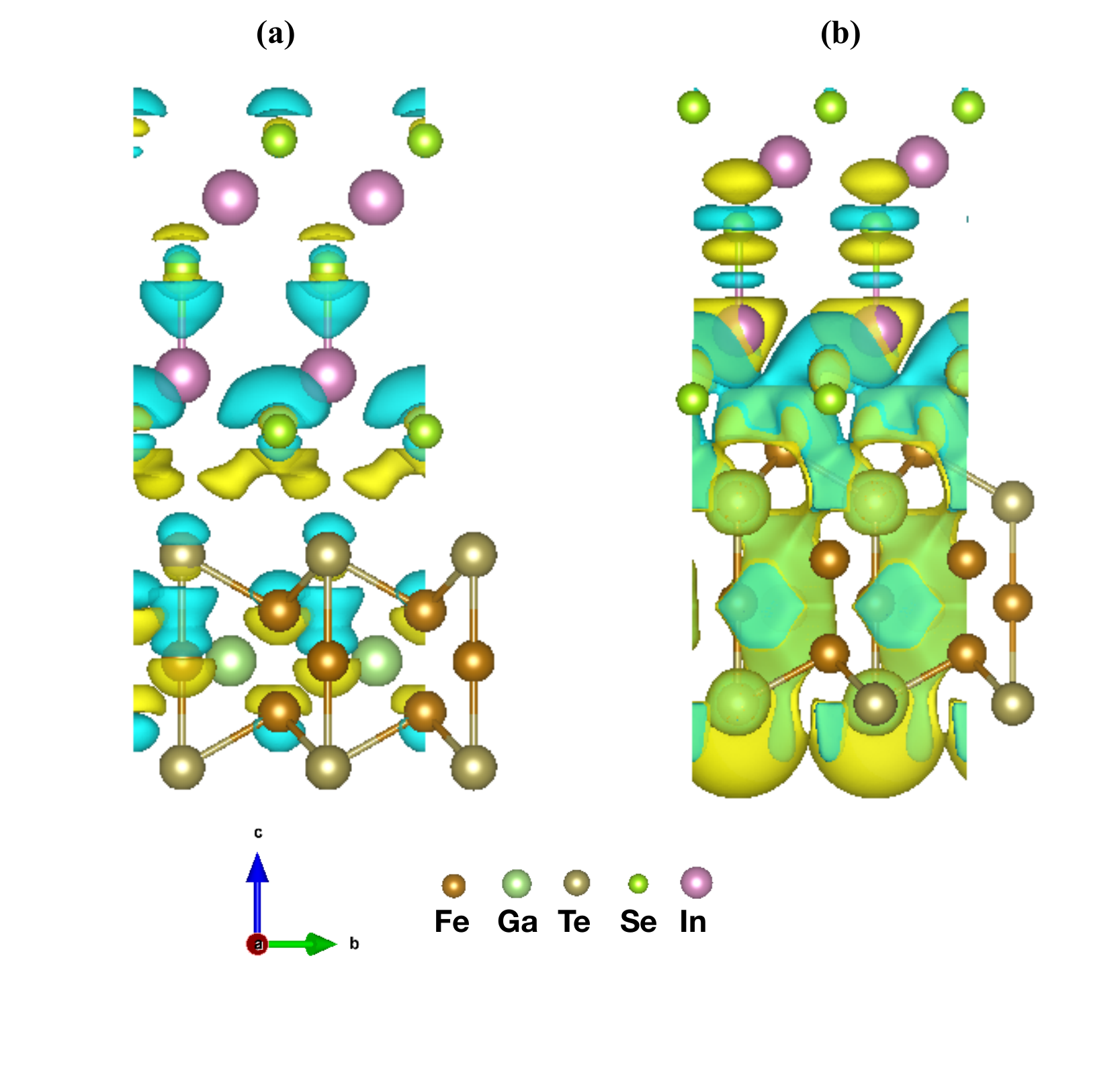}
    \caption{Differential charge density (DCD) illustrating charge redistribution in (a) Fe$_3$GaTe$_2$/In$_2$Se$_3$ heterostructure and (b) Fe-intercalated Fe$_3$GaTe$_2$/In$_2$Se$_3$ heterostructure, both for the up polarization state of In$_2$Se$_3$. An isosurface value of 0.0005~e/Bohr$^3$ is used. Yellow and cyan contours represent regions of charge accumulation and depletion, respectively, arising from the interfacial interaction between the two layers.}
    \label{DCD}
\end{figure}

As shown in Fig.~\ref{DCD}, the differential charge density (DCD) plot reveals the trend of charge redistribution in both the pristine heterostructure and the Fe-intercalated heterostructure. DCD is the difference in charge densities of the heterostructure and individual isolated layers. It shows regions of charge accumulation and depletion as the heterostructure is formed. 
In the Fe-intercalated heterostructure, a higher accumulation of charge in the FGaT layer, compared to the pristine case, is responsible for the increased polarization in the upward direction. Moreover, a significant amount of charge density at the interlayer Fe site is clearly visible, indicating strong interfacial hybridization.

Together, these results establish a promising platform for electrically controllable magnetic anisotropy, driven by interlayer hybridization and polarization switching as key mechanisms for next-generation magnetoelectric and spintronic devices.

\subsection*{Explaining the origin of anisotropy }

\begin{figure*}
    \centering
    \includegraphics[width=0.7\textwidth]{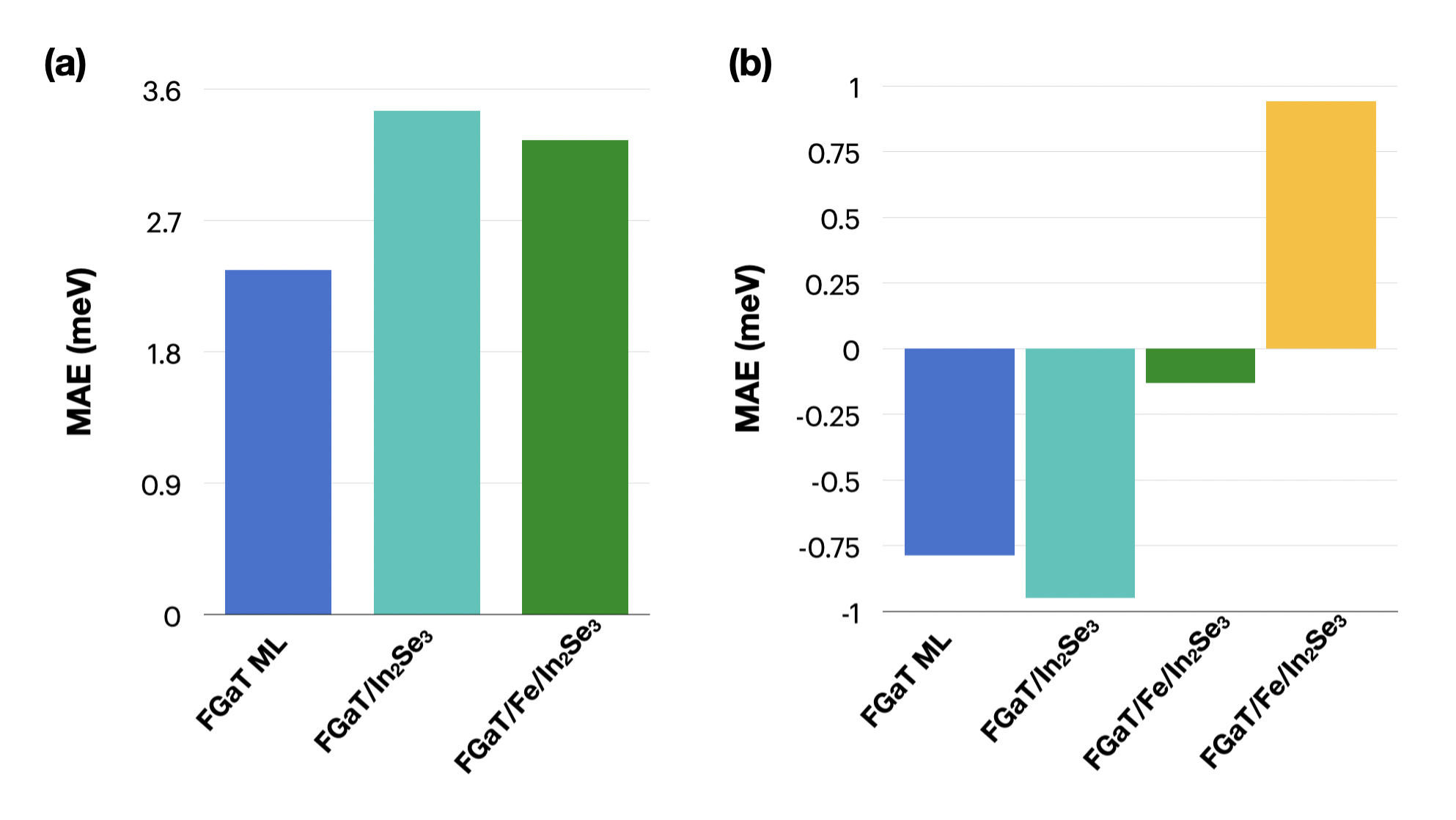}
    \caption{Atom resolved MAE contributions for the three systems: monolayer FGaT, pristine FGaT/In$_2$Se$_3$ heterostructure, and Fe-intercalated FGaT/In$_2$Se$_3$ heterostructure. (a) MAE contribution of Te atoms (b) MAE contribution of Fe atoms. The yellow bar is the contribution from the intercalated Fe atom.   }
    \label{atom-resolved-mae}
\end{figure*}

To elucidate the origin of large PMA, we analyzed the atom-resolved contributions to the magnetic anisotropy energy (MAE) for each system, as summarized in Fig.~\ref{atom-resolved-mae}. The contributions from Te atoms are presented in panel (a) of the figure, while those from Fe atoms are shown in panel (b). The dominant contribution to the MAE arises from Te atoms, which play a crucial role in establishing the PMA. In monolayer FGaT, Te atoms contribute 2.362 meV, whereas Fe atoms contribute –0.785 meV, indicating that Te atoms favor out-of-plane (easy-axis) anisotropy, while Fe atoms prefer in-plane (easy-plane) anisotropy.

In the FGaT/In$_2$Se$_3$ heterostructure, the contribution from Te atoms increases by 1.088 meV, whereas the Fe contribution becomes slightly more negative by 0.164 meV. The net increase in MAE is therefore primarily driven by the enhanced Te contribution. In the Fe-intercalated heterostructure, the Te contribution remains nearly unchanged compared to the pristine heterostructure. However, the intrinsic Fe contribution is significantly reduced in magnitude to –0.131 meV, and notably, the intercalated Fe atom contributes positively with 0.942 meV. This sign reversal and enhanced contribution from the intercalated Fe atom account for the further increase in MAE in the Fe-intercalated system ~\cite{perdew1996generalized}.


The atom-resolved contributions to the MAE can be further understood by analyzing the underlying electronic structure of each system. The MAE originates from spin-orbit coupling (SOC), which can be viewed as a relativistic perturbation to the non-relativistic Hamiltonian since the SOC energy scale ($\sim$meV) is much smaller than the typical electronic band energies ($~\sim$eV). The first-order energy correction vanishes, and thus the dominant contribution to MAE arises from second-order perturbation theory~\cite{Gong2019-ct, PertMAE},


\begin{align}
\Delta E^{\mathrm{SOC}}(s) =\ & 
 \lambda^2 \sum_{o,u,\sigma} 
\frac{|\langle o , \sigma | \mathbf{L}_z | u, \sigma \rangle|^2 
- |\langle o, \sigma | \mathbf{L}_x | u, \sigma \rangle|^2}{E_u - E_o} \nonumber \\
& 
+ \lambda^2 \sum_{\substack{o,u \\ \sigma \neq \sigma'}} 
\frac{|\langle o, \sigma | \mathbf{L}_x | u, \sigma' \rangle|^2 
- |\langle o, \sigma | \mathbf{L}_z | u, \sigma' \rangle|^2}{E_u - E_o}
\label{eq:PMA}
\end{align}

Here, $|o, \sigma\rangle$ and $|u, \sigma\rangle$ denote the occupied and unoccupied electronic states with spin $\sigma$, and $E_o$ and $E_u$ are their corresponding eigenenergies. The first term in Eq.~\eqref{eq:PMA} accounts for spin-conserving transitions, while the second term represents spin-flip transitions. Both are mediated by SOC through angular momentum matrix elements.

Importantly, the dominant contributions to the MAE originate from SOC-induced coupling between states near the Fermi level, due to the energy denominator $(E_u - E_o)$ in Eq.~\eqref{eq:PMA}. The sign and magnitude of each contribution are determined by the nature of the transition and the angular momentum channel involved. For spin-conserving transitions, matrix elements involving $L_z$ enhance out-of-plane (easy-axis) anisotropy, while those involving $L_x$ favor in-plane (easy-plane) anisotropy. Conversely, for spin-flip transitions, $L_z$ contributes negatively and $L_x$ positively to the anisotropy. As a result, a large density of states (DOS) near the Fermi level can significantly influence both the magnitude and preferred direction of the MAE.

We next focus on the Fe atoms to analyze the orbital-resolved origin of MAE across all systems. The nonzero SOC matrix elements relevant to these transitions include
$\langle d_{yz} | L_x | d_{z^2} \rangle$,
$\langle d_{xy} | L_x | d_{xz} \rangle$,
$\langle d_{yz} | L_x | d_{x^2 - y^2} \rangle$,
$\langle d_{xy} | L_z | d_{x^2 - y^2} \rangle$, and
$\langle d_{yz} | L_z | d_{xz} \rangle$.

We first analyze the orbital-resolved MAE contributions shown in Fig.~\ref{atom-resolved-mae}(a), (c), and (e). Among all the transitions, the most significant matrix element is $\langle d_{xy} | L_z | d_{x^2 - y^2} \rangle$. In monolayer FGaT, this transition contributes negatively to the MAE, with a value of $-0.19$ meV, and similarly contributes $-0.26$ meV in the pristine FGaT/In$_2$Se$_3$ heterostructure. However, in the Fe-intercalated heterostructure, it contributes a large positive value of 0.98 meV, which is the primary reason for the enhanced PMA in this system.

\begin{figure*}
    \centering
    \includegraphics[width=1.0\textwidth]{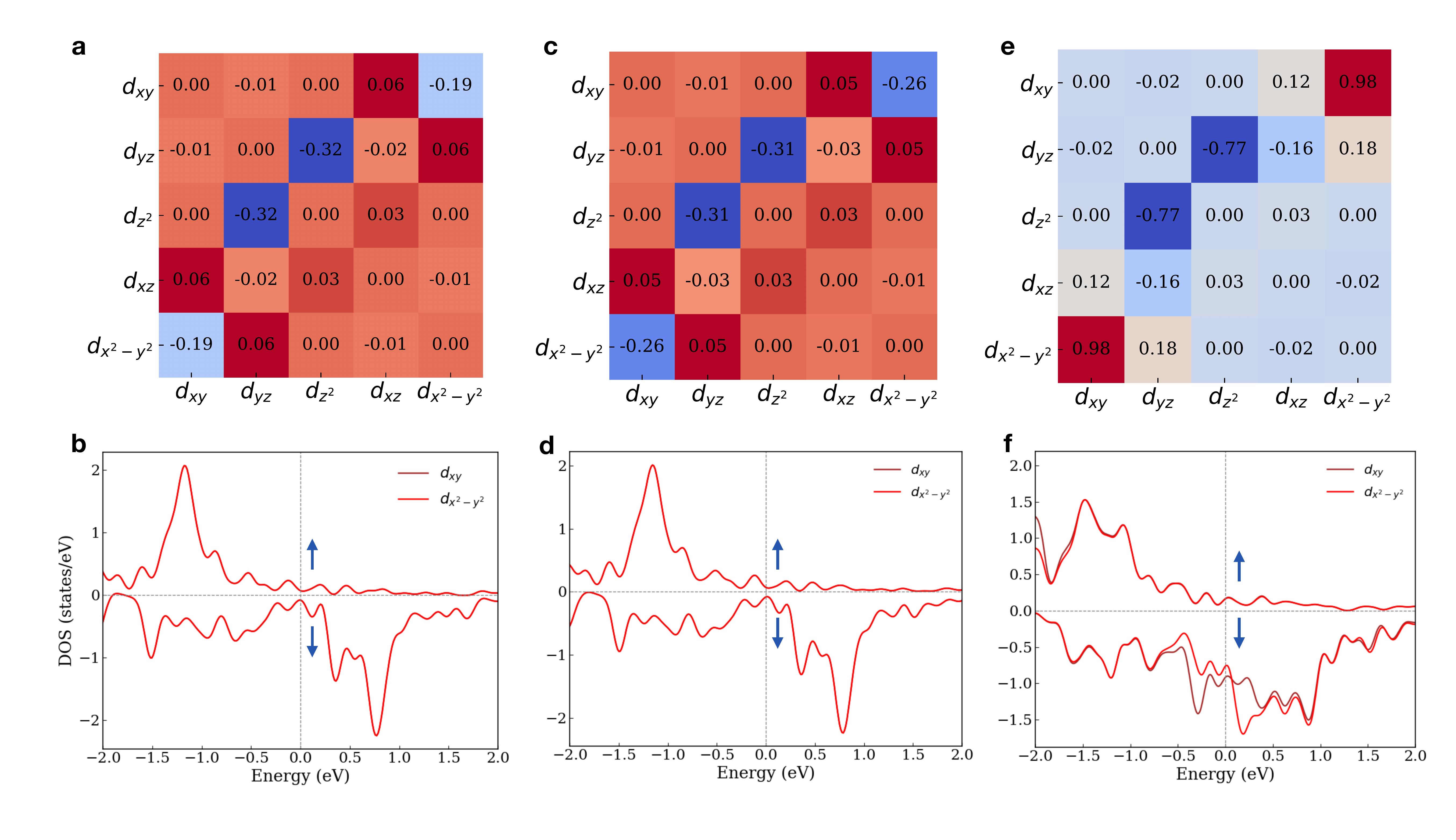}
    \caption{The heat map of net MAE contribution of Fe atoms in units of meV, calculated from the differences of spin–orbit coupling matrix elements, and the corresponding density of states (DOS) of Fe $d_{xy}$ and $d_{x^2-y^2}$ orbitals in (a)-(b) monolayer FGaT, (c)-(d) pristine FGaT/\text{In}$_2$\text{Se}$_3$ heterostructure, and (e)-(f) Fe-intercalated FGaT/\text{In}$_2$\text{Se}$_3$ heterostructure. }
    \label{DOS_SOC}
\end{figure*}

These trends can be directly correlated with the orbital-projected density of states (DOS), shown in Fig.~\ref{DOS_SOC}(b), (d), and (f). In both monolayer FGaT and the pristine heterostructure, the $d_{x^2 - y^2}$ and $d_{xy}$ orbitals exhibit identical DOS profiles, and the dominant SOC-induced transitions involve opposite spin channels (i.e., spin-up occupied to spin-down unoccupied states). According to the perturbative expression for MAE, such spin-flip transitions involving $L_z$ yield negative contributions, consistent with the observed values.

In contrast, in the Fe-intercalated heterostructure, a substantial increase in the DOS is observed in the spin-down channel near the Fermi level, particularly at the Fermi level itself. This results in dominant spin-conserving transitions within the same spin channel (spin-down occupied to spin-down unoccupied), which, through the $L_z$ operator, contribute positively to the MAE. The large DOS near the Fermi level further enhances the magnitude of this contribution, thereby explaining the significantly increased PMA in the Fe-intercalated system.



\section{Conclusions}
In summary, we have studied a two-dimensional van der Waals bilayer heterostructure composed of Fe$_3$GaTe$_2$ (FGaT) and In$_2$Se$_3$. FGaT is a known room-temperature ferromagnet, although it exhibits relatively low PMA. Our investigation demonstrates that interfacial hybridization and Fe self-intercalation in the heterostructure act as effective mechanisms to enhance the PMA of this high-$T_c$ ferromagnetic material.
Additionally, we observe significant tunability of electric polarization in the system, strongly influenced by the interaction between the In$_2$Se$_3$ polarization state and the adjacent FGaT layer. The interlayer coupling between the two materials leads to substantial charge redistribution across the interface, as clearly evidenced by the differential charge density plots.

This work highlights a promising strategy for tuning magnetic and ferroelectric properties in 2D van der Waals heterostructures, offering significant implications for the design of next-generation spintronic and multifunctional devices.

\clearpage 
\bibliographystyle{apsrev4-2}
\bibliography{ref.bib}
\end{document}